


\input harvmac.tex
\input tables.tex
\noblackbox

\def\pmb#1{\setbox0=\hbox{#1}%
\kern-.025em\copy0\kern-\wd0
\kern.05em\copy0\kern-\wd0
\kern-.025em\raise.0433em\box0 }

\def\xhat{{\bf \hat{x}}}
\def\yhat{{\bf \hat{y}}}
\def\zhat{{\bf \hat{z}}}
\def\rvec{{\bf r}}

\def\rpvec{{\bf r}_{\perp}}

\def\lp{\lambda_{ab}}

\def\epst{\tilde{\epsilon}_1}
\def\loc{l_{\perp}}


\Title{}{\vbox{\centerline{Twin-Boundary Pinning}
\vskip2pt\centerline{of Superconducting Vortex Arrays}}}

\baselineskip= 16pt plus 2pt minus 1pt\centerline{M. CRISTINA MARCHETTI
\footnote{\dag}
{E-mail: cris@npac.syr.edu}}
\centerline{Physics Department}
\centerline{Syracuse University}
\centerline{Syracuse, NY 13244}
\medskip
\centerline{and}
\medskip
\centerline{VALERII M. VINOKUR
\footnote{*}
{E-mail:valerii\_vinokur@qmgate.anl.gov}}
\centerline{Material Science Division}
\centerline{Argonne National Laboratory}
\centerline{Argonne, IL 60439}
\vskip .1in

We discuss the low-temperature dynamics of magnetic flux lines in
high-temperature superconductors in the presence of a family of
parallel twin planes that contain the $c$ axis. A current applied along the
twin planes drives flux motion in the direction transverse to the
planes and acts like an electric field applied to {\it one-dimensional}
carriers in disordered semiconductors.
As in flux arrays with columnar pins, there is a regime where the dynamics
is dominated by superkink excitations
that correspond to Mott variable
range hopping (VRH) of carriers. In one dimension, however,
rare events, such as large regions void of twin planes, can impede
VRH and dominate transport in samples that are sufficiently long
in the direction of flux motion. In short samples
rare regions can be responsible for mesoscospic effects.
The phase boundaries separating various transport regimes
are discussed.
The effect of tilting the applied field out of the twin planes is also
considered. In this case the resistivity from flux motion is found to depend
strongly on the tilt angle.

\Date{5/94}

\newsec{Introduction}

The static and dynamical properties of magnetic flux lines
in copper-oxide superconductors are strongly affected by
pinning by point, linear and planar disorder
\nref\tora{For a review and an extensive list of references,
see G. Blatter, M.V. Feigel'man, V.B. Geshkenbein, A.I. Larkin,
amd V.M. Vinokur, Rev. Mod. Phys., to appear.}\refs{\tora}.
Twin boundaries are an example of planar disorder that is generally present in
superconducting $YBa_2Cu_3O_{7-x}$ and $La_2CuO_4$, where
they are needed to accomodate the strains produced by a crystallografic
tetragonal-to-orthorhombic transition. Twins most often occurs in
two orthogonal families of lamellae forming a mosaic \nref\mitchell{T. Roy
and T.E. Mitchell, Phil. Mag. A {\bf 63}, 225 (1991);
T.E. Mitchell and J.P. Hirth, Acta Metall. Mater. {\bf 39}, 1711 (1991).}
\refs{\mitchell}. It is also
possible to prepare samples that contain a single family of parallel
twin planes \nref\crabtree{G.W. Crabtree,
W.K. Kwok, U. Welp, J. Downey, S. Flesher, K.G. Vandervoort
and J.Z. Lin, Physica C {\bf 185-189}, 282 (1991).}\nref\kwoktwo{W.K. Kwok,
S. Flesher, U. Welp, V.M. Vinokur, J. Downey,
G.W. Crabtree, and M.M. Miller, Phys. Rev. Lett. {\bf 69}, 3370 (1992);
and S. Flesher, W.K. Kwok, U. Welp, V.M. Vinokur, M.K. Smith, J. Downey,
and G.W. Crabtree, Phys. Rev. {\bf B47}, 14448 (1993).}\refs{\crabtree ,
\kwoktwo}.
Columnar or linear defects can be produced by
bombardment of the crystal with energetic heavy ions \nref\konc{M. Konczykowski
et al., Phys. Rev. B {\bf 44}, 7167 (1991).}\refs{\konc}
or are embodied
in forests of screw dislocations parallel to the crystal growth
direction \nref\hawley{M. Hawley, I.D. Raistrick, J.G. Beery and R.J.
Houlton, Science {\bf 251}, 1587 (1991); C. Gerber, D. Anselmetti, J.G.
Bednorz, J. Manhort and D.G. Schlom, Nature {\bf 350}, 279
(1991).}\refs{\hawley}.
Both twins \nref\kwokone{W.K. Kwok,
U. Welp, G.W. Crabtree, K.G. Vandervoort, R. Hulscher and J.Z. Liu,
Phys. Rev. Lett. {\bf 64}, 966 (1990).}\refs{\kwokone ,\crabtree, \kwoktwo}
and columnar defects \nref\civale{L. Civale, A.D. Marwick,
T.K. Worthington,
M.A. Kirk, J.R. Thompson, L. Krusin-Elbaum, Y. Sun, J.R. Clem,
and F. Holtzberg, Phys. Rev. Lett. {\bf 67}, 648 (1991); R.C. Budhani,
M. Suenaga and S.H. Liou, Phys. Rev. Lett. {\bf 69}, 3816 (1992).}
\refs{\civale} constitute examples of macroscopic correlated
disorder that can be responsible for a sharp decrease
in the resistivity for specific field orientations.

Extensive investigations of twin-boundary pinning have been carried out
by Kwok and coworkers\refs{\kwoktwo}. These authors studied a variety of YBCO
single
crystal samples containing single families of parallel twins lying
in planes spanned by the $c$ axis, with
spacings ranging from microns down to several hundred Angstroms.
As indicated by earlier decoration experiments\nref\dolan{G.J. Dolan,
G.V. Chandrashekhar, T.R. Dinger,
C. Feild and F. Holtzberg, Phys. Rev. Lett. {\bf 62}, 827 (1989).}
\refs{\dolan}, the order parameter
is suppressed at a twin boundary at low temperatures and the twin attracts
or pins the vortices.
Transport experiments show clear evidence of strong
twin-boundary pinning even in the flux liquid phase \refs{\kwokone}.
The pinning is most effective when the
external field is applied along the $c$ axis and the driving current
lies in the $ab$ plane and is parallel to the plane of the twins, resulting in
a Lorentz force normal to the twin planes (Fig. 1a).
In this case twin planes act like
strong pinning centers and
the linear resistivity as a function of temperature
drops sharply at a characteristic
temperature that marks the onset of twin-boundary pinning
\refs{\kwokone}.
In samples with only a few widely spaced twin planes this drop in
the resistivity is followed by an abrupt shoulder at a lower temperature,
corresponding to the first order freezing transition
into an Abrikosov lattice. The abrupt shoulder is not observed in
heavily twinned samples where the freezing transition is
apparently suppressed by disorder.
Additional experimental evidence for twin-boundary pinning comes from the
observation of a sharp downward dip in the
resistivity as a function of angle as the external
field is rotated out of the twin plane through the $c$
direction \refs{\kwoktwo} (see Fig. 1b).
This strong angular dependence is a clear signature of anisotropic pinning
by twin planes since
point disorder in the form of oxygen vacancies,
while certainly present in all these samples, can only yield
a weak dependence of the resistivity on the tilting angle.

Another experimental probe of flux-line dynamics that has been recently used to
study the effect of correlated disorder is
real-time imaging of
flux profiles \nref\duran{C. Dur\'an,
P.L. Gammel, R. Wolfe,
V.J. Fratello, D.J. Bishop, J.P. Rice and D.M. Ginsberg, Nature {\bf 357}, 474
(1992).}\nref\vlasko{V.K.
Vlasko-Vlasov,
L.A. Dorosinskii, A.A. Polyanskii, V.I. Nikitenko, U. Welp, B.W. Veal
and G.W. Crabtree, Phys. Rev. Lett. {\bf 72}, 3246 (1994).}\refs{\duran ,
\vlasko}.
Imaging experiments of the field penetrating into single crystals with
families of twins lying in planes spanned by the $c$ axis (for fields
directed
along the $c$ axis) show strong pinning by twin planes for flux motion in
the direction normal to the planes,
confirming the strong twin-boundary pinning seen by
transport measurements for Lorentz forces applied normal to the twin planes.
Imaging of flux motion
along the twin planes have, however, given contradictory results.
Dur\'an and coworkers \refs{\duran} argued that flux penetrates more easily in
the twin regions
than in the channels between the twin planes, in apparent
contrast with the observation of a lowered flux-flow resistivity
by twin-boundary pinning by the Argonne group for the parallel geometry,
where the Lorentz force is applied along the plane of the twins
\nref\argeug{W.K. Kwok, J. Fendrich, S. Fleshler, U. Welp, J. Downey,
G.W. Crabtree, and J. Giapintzakis, in {\it Proceedings of the XX
International Conference on Low Temperature Physics}, August 4-11, 1993.}
\refs{\argeug}.
A more recent imaging experiment
by Vlasko-Vlasov and collaborators \refs{\vlasko}
shows, however, that the twin planes
act as
planar pinning barriers for all directions of flux flow, giving rise to
guided vortex motion.
In this paper we are only concerned
with transport with flux motion in the direction transverse to the twin
planes. For this geometry all experimental probes of flux dynamics
confirm that twin planes provide attractive potential wells
for the vortices. A detailed theoretical understanding
of the imaging experiments and their relationship to transport
transport in the parallel geometry remains, however,
an open question.

The static and dynamical properties of flux line assemblies in the presence of
a random array of {\it columnar pins} have been studied in detail by mapping
the physics of magnetic flux lines onto the problem of localization
of quantum mechanical bosons in {\it two dimensions}
\nref\nvlong{D.R. Nelson and V.M. Vinokur, Phys. Rev. Lett. {\bf 68}, 2398
(1992); D.R. Nelson and V.M. Vinokur, Phys. Rev. B {\bf 48}, 13060
(1993).}
\refs{\nvlong}. This mapping exploits methods
developed to understand the behavior of $He^4$ films on disordered
substrates \nref\grinstein{M.P.A. Fisher, P.B. Weichman, G. Grinstein,
and D. S. Fisher, Phys. Rev. B {\bf 40}, 546 (1989).}\refs{\grinstein}
and to decribe electronic transport in disordered superconductors
\nref\shklovskii{B.I. Shklovskii and A.L. Efros, {\it Electronic Properties
of Doped Semiconductors} (Springer-Verlagh, New York,
1984).}\refs{\shklovskii}.
At low temperatures there is a ``Bose glass'' phase, with flux lines localized
on columnar pins, separated by a phase transition from an entangled flux
liquid of delocalized lines. In the Bose glass phase the linear resistivity
vanishes and the current-voltage characteristics are nonlinear. Transport in
this
regime closely resembles the variable-range hopping (VRH) of electrons in
disordered semiconductors in two dimensions \refs{\shklovskii}.

In this paper we employ similar methods to study flux-line dynamics
in the presence
of a {\it single family of parallel twin boundaries} lying in planes containing
the $c$ axis. For $\vec{H}\parallel\hat{c}$, the flux lines are localized
by the pinning potential in the direction normal to the planes.
At low temperatures, when the average vortex spacing $a_0\approx(\phi_0/B)$
exceeds the average distance $d$ between twin planes,
all flux lines are localized on the twins,
progressively ``filling'' the planar pins as the field is
increased.
We only consider vortex arrays that are sufficiently dilute
($B<<B_f\approx \phi_0/d^2$) that transport is controlled by single-vortex
creep. This regime is experimentally relevant and has in fact been probed
in flux arrays with columnar defects \nref\konc{}\refs{\konc}.
Flux motion in this regime is dominated by thermally activated jumps of
the vortices
over the relevant pinning energy barriers ${\cal U}(L,J)$, yielding
a resistivity $\rho={\cal E}/J$, given by \refs{\tora},
\eqn\taff{\rho(T)\approx \rho_0e^{-{\cal U}(L,J)/T},}
where $\rho_0$ is a characteristic flux-flow resistivity.
Here we focus
on flux motion transverse to the twin planes at low fields and temperatures,
which resembles the hopping of
electrons in {\it one-dimensional} disordered superconductors.
The energy barriers ${\cal U}$ corresponding to the various low-lying
excitations that can contribute to transport are evaluated and
are summarized in Table 1.
Typical phase diagrams in the $(J,L)$ plane displaying the regions
where the different transport mechanisms dominate are shown in Fig. 2.
There is a characteristic current scale $J_L\sim 1/L$, where $L$
is the sample thickness in the field direction,
that separates the regions of linear and nonlinear
response. As $L\rightarrow\infty$ the response is always nonlinear.
For large enough current the dominant excitations are the half-loop
configurations shown in Fig. 3a, of transverse width smaller than the average
separation $d$ between twin planes. For currents below
$J_1$ the width of a typical half-loop excitation exceeds the mean distance
between pins.
In this case standard VRH arguments \refs{\shklovskii , \nvlong},
where an electron hops larger and larger spatial distances to
find ``good'' traps of low energy (Fig. 3c), suggest a nonlinear
current-voltage
characteristic $V\sim \exp[-(E_k/T)(J_0/J)^{1/2}]$ in the localized phase,
where $E_k$ and $J_0$ are characteristic energy and current scales
given below. The phase diagrams shown in Fig. 2 are qualitatively similar
to the phase diagrams one would obtain for vortex arrays in the presence
of columnar pins. In that case one finds a nonlinear current-voltage
characteristic at low currents typical of VRH in two dimensions,
with $V\sim \exp[-(E_k/T)(J_0/J)^{1/3}]$.

The main difference between flux motion in the presence of columnar defects
and flux motion transverse
to an array of parallel twin planes is that in the latter case the low
temperature
transport is {\it one-dimensional}.
In one dimension VRH can be impeded
by the presence of large rare regions
void of energetically favorable pins \nref\kurk{J. Kurkijarvi,
Phys. Rev. B {\bf 8}, 922 (1973).}\nref\ruzin{M.E. Raikh and I.M. Ruzin,
Zh. Eksp. Teor. Fiz. {\bf 95}, 1113 (1989) [Sov. Phys. JETP {\bf 68}(3), 642
(1989)].}
\refs{\kurk , \ruzin}.
These rare regions free of localized states are exponentially rare, but
have a very large resistance and can suppress VRH or even dominate
transport (Fig. 4) in one
dimension since the vortices cannot get around them.
For samples that are sufficiently long in the direction of flux-line
motion so that they contain a large number of such rare regions,
this mechanism yields a nonlinear current-voltage characteristic of the
form $V\sim(T/E_k)(J/J_0)^{1/2}\exp[-(E_k/T)^2(J_0/J)]$.
Shorter samples will typically contain only a few
rare regions and these will determine the sample's resistance.
In sufficiently short samples there will be a spread of values of the
resistance between different samples resulting in reproducible
sample-to-sample resistance fluctuations.

Finally, the model presented here is also relevant to the dynamics of
Josephson vortices in artificially structured ``giant'' Josephson junctions.
Consider for instance a planar junction
of in-plane dimensions large
compared to the
Josephson penetration length at the contact of two superconductors
(the plane of the junction is the $xz$ plane).
A magnetic field applied in the plane of the junction
(say, in the $z$ direction)
penetrates into the junction
as a chain of Josephson vortices which lie in the contact plane
\nref\scalapino{C.S. Owen and D.J. Scalapino, Phys. Rev. {\bf 164},
538 (1967).}\nref\vvkosh{V.M. Vinokur and A.E. Koshelev, Sov. Phys. JETP
{\bf 70}, 547 (1990).}\refs{\scalapino , \vvkosh}.
The intervortex  spacing along the $x$ direction is determined by the
strength of the applied field. The vortices are localized in
the plane of the junction
and form therefore a $(1+1)$-dimensional vortex array in this
plane.
If the junction is not uniform, the vortices are pinned
independently at low fields, as indicated by the fact that the critical
current does not depend on magnetic field.
Additional defects can be artificially introduced in the junction.
The junction can be artificially structured by the
introduction of an array of defects along the $x$ direction spanning the
field axis ($z$) and the junction thickness ($y$).
At low temperature vortex motion along the junction plane will then
occur via thermally activated jumps between these defects and will be
described by the one-dimensional tight-binding model introduced below.

In Section 2 we first review the simple model of interacting flux lines
in the presence of correlated disorder introduced by
Nelson \nref\drnrev{D.R.
Nelson, in Proceedings of the Los Alamos
Symposium {\it Phenomenology and Applications of High-Temperature
Superconductors}, K.S. Bedell et al., eds. (Addison-Wesley, 1991),
and references therein.}\refs{\drnrev}
and by Nelson and Vinokur \refs{\nvlong}. The analogy with quantum mechanics
of {\it two-dimensional} bosons and the reduction of the low temperature
dynamics transverse to an array of parallel twin planes to a tight
binding model in {\it one dimension} are then discussed.
In Section 3 we estimate the pinning energy barriers associated with the
low-lying excitations from the ground state and the corresponding contributions
to the resistivity. The phase boundaries
separating the regions of the $(L,J)$ plane where the various contributions
dominate are also discussed. A brief summary of these results has been
presented elsewhere \nref\mcmvv{M.C. Marchetti and V.M. Vinokur,
Phys. Rev. Lett. {\bf 72}, 3409 (1994).}\refs{\mcmvv}.
In section 4 we consider the case where the external field is tilted at
an angle $\theta$ away from the $c$ axis and out of the twin planes.
The resistivity displays a strong angular dependence with a sharp
downward dip at $\theta=0$.
Finally, in Section 5 we discuss the role
of rare fluctuations in this one dimensional geometry.

\newsec{Vortex Free Energy and Tight-binding Model}

We are interested in transport at low fields and temperatures
where each flux line is localized on one or more twin planes.
In this regime dominated by single-vortex dynamics a detailed
description of transport can be developed.
Our starting point is a model-free energy
for flux lines in a sample of thickness $L$ in the presence of a
family of parallel twin planes \refs{\nvlong}.
The field is along the $c$ axis, chosen
as the $z$ direction, and the flux lines are parametrized by their
trajectories $\{\rvec_i(z)\}$ as they traverse the sample. The twin
boundaries are parallel to the $zx$ plane. The model free-energy
for a single flux line at $(\rvec_1(z),z)$ is
given by
\eqn\freeone{F_1=\int_0^L dz \bigg[{\epst\over 2}
  \Big|{d\rvec_1(z)\over dz}\Big|^2
  +V_D(y_1(z))\bigg],}
with
\eqn\dispot{V_D(y)=\sum_{k=1}^M V_1(|y-Y_k|).}
Here $V_D$ is the random potential arising from a set of $M$ $x$- and
$z$-independent pinning potentials $V_1(|y-Y_k|)$ centered at the locations
$\{Y_k\}$ of the twin planes. The first term on the
right hand side of Eq. \freeone\ is the first term
in a small angle expansion of the elastic energy of a nearly straight
vortex line, with $\epst\approx (M_{\perp}/M_z)\epsilon_0\ln(\lp/\xi_{ab})$
the tilt modulus and $\lp$ and $\xi_{ab}$ the penetration
and the coherence lengths in the $ab$ plane, respectively.
The effective mass ratio $M_{\perp}/M_z<<1$ incorporates
the material anisotropy and $\epsilon_0\approx (\phi_0/4\pi\lp)^2$
is a characteristic energy scale.
For simplicity we model $V_D(y)$ as an array of identical one-dimensional
square potential wells of depth $U_0$, width $2b_0$ and average spacing $d$,
passing completely through the sample in the $x$ and $z$ directions
\nref\blatter{G. Blatter, J. Rhyner and V.M. Vinokur, Phys. Rev. B
{\bf 43}, 7826 (1991).}\refs{\blatter}.
Assuming the potential wells are centered at uniformly distributed
random positions $\{Y_k\}$ and $b_0<<d$, we find
$\overline{V_D}\approx U_0\Big({2b_0\over d}\Big)$, while the
random potential fluctuations $\delta V_D(y)=V_D(y)-\overline{V_D}$
satisfy
\eqn\potfluct{\overline{\delta V_D(y)\delta V_D(y')}=\Delta\delta (y-y'),}
with $\Delta\approx U_0^2{(2b_0)^2\over d}[1+{\cal O}(2b_0/d)]$.
The interaction between vortex lines and a twin boundary has been studied
by Geshkenbein in the context of the Ginzburg-Landau theory \nref\gesh{V.B.
Geshkenbein, Sov. Phys. JETP {\bf 67}, 2166 (1988).}\refs{\gesh}.

In this paper we assume that the twin planes are randomly
located along the $y$ direction and that
the separations between neighboring twins are Poisson-distributed
(see also Section 5 below).
This appears to be the case in some of the samples employed by the Argonne
group \nref\kwokpriv{W.K. Kwok, private communication.}\refs{\kwokpriv}.
On the other hand, the twin structures that form naturally
in YBCO to accomodate the strains arising from a tetragonal-to-orthorhombic
transformation, which take place around $1000K$ as a result of oxygen
vacancy ordering, are often quite different \refs{\mitchell}.
They consists of lamellae or colonies of parallel twins oriented in either
the $(110)$ or $(1\overline{1}0)$ directions. Orthogonal twin colonies
form a mosaic-type structure, containing colonies of various size.
The colony size scales with the square of the average twin spacing $d$,
while the latter remains rather uniform within a given colony\refs{\mitchell}.
There is a repulsive interaction between the twin planes of a given colony
arising from the stress produced by one twin plane in the region of another.
This interaction leads to the regular spacing of the twins within a colony
which resembles an approximately regular
one-dimensional lattice of twin planes. Vortex dynamics in the presence
of such a twin structure will not be described by the model presented
in this paper. On scales shorter than the typical twin colony
size vortices are pinned by a regular array of planar defects,
while on scales larger than the colony size the theory developed to
describe vortex dynamics in the presence of columnar defects should apply.
In contrast, the twin structures observed in the samples used by the Argonne
group
consist of a single colony of parallel twins with large variations in the twin
spacing. A twin structure of this type may arise if the sample is annealed
and the twins ``fall out of equilibrium'' arranging themselves in
one-dimensional liquid-like structure within a given colony.

The free energy for an assembly of $N$ flux lines
is given by
\eqn\freemany{F_{(N)}=\sum_{i=1}^N F_i + {1\over 2}\sum_{i\not= j}
 \int_0^L V(|\rvec_i(z)-\rvec_j(z)|) dz,}
where $V(r)$ is the pair interaction potential, assumed local in $z$.
It can be shown \refs{\nvlong} that both higher order terms in the
small angle expansion of the
elastic energy and nonlocality in $z$ in the pair interaction
are negligible provided $|d\rvec_i/dz|^2<<M_z/M_{\perp}$ for the most
important vortex configurations. In the following we will
simply use the form of the pair interaction for nearly
straight flux lines,
\eqn\pairint{V(r)\approx 2\epsilon_0 \big[K_0(r/\lambda_{ab})
   -K_0(r/\xi_{ab})\big],}
where $K_0(x)$ is a modified Bessel function.

At low temperatures, when the average vortex spacing $a_0\approx(\phi_0/B)$
exceeds the average distance between twin planes, all flux lines are localized
on the twins, progressively ``filling'' the planar pins as the field is
increased.
Any real sample will, however, also contain point defects, which are known to
promote flux-line wandering.
Recent analytical and numerical work has shown that
in the case of planar pins a single flux line remains localized on the pinning
plane even in
the presence of additional weak point disorder in the bulk, which is
always present in real samples
\nref\balents{L. Balents and M. Kardar, Europhys. Lett. {\bf 23}, 503 (1993).}
\refs{\balents}.
The stability of the localized Bose glass phase in $1+1$ dimensions - the
case relevant to the model considered here - in the presence of point
disorder has been studied recently by Hwa et al. \nref\hwa{T. Hwa, D.R. Nelson
and V.M. Vinokur, Phys. Rev. B {\bf 48}, 1167 (1993).}\refs{\hwa}.
These authors considered a model where flux lines directed along the $z$
direction and confined to the $zy$ plane are pinned by the competing
action of randomly distributed linear defects spanning the plane in the $z$
direction and point defects described by a random potential
with variance $\Delta_0$. They showed that in $1+1$ dimensions the
low temperature phase is of the Bose glass type with flux lines
localized on the linear pins when point
disorder is weak. The localized phase is marginally unstable to point disorder,
but only beyond an astronomically large crossover length scale.
Point disorder will in general lower the energy barriers associated
with the various low-lying excitations discussed here. It does not, however,
have a significant effect on the energy barriers in the rigid flow
(or half-loop) regime where ${\cal U}_{rf}\sim L$ (or ${\cal U}_{hl}\sim 1/J$).
This is because the energy gain $\delta F_{\Delta}$ associated with
the pinning of a fluctuation
of length $L$ by point defects only grows as $L^{1/2}$
(or as $J^{-1/2}$ in the nonlinear regime), with
$\delta F_{\Delta}\sim (\Delta_0 L)^{1/2}$.
Sufficiently strong point disorder can, however, lower considerably
the barriers for variable range hopping \`a la Mott
since both the barrier ${\cal U}_{Mott}$ and $\delta F_{\Delta}$
grow like $L^{1/2}$. In this case
to assess whether point or correlated disorder dominates one needs to
compare quantitatively the relative strengths of these two types of crystal
defects.
This comparison
involves unknown parameters and is beyond the scope of the present paper.

The problem of one flux line localized near a single twin plane
has been studied by Nelson by exploiting the mapping of the statistical
mechanics of magnetic flux lines onto the quantum mechanics of
two-dimensional bosons \refs{\drnrev}.
At zero temperature
the twin provides a binding energy $U_0$ per unit length for trapping the
flux line. In the presence of thermal fluctuations $U_0$ is replaced by a
smaller binding free energy per unit length $U(T)$, to account for the entropy
lost by confining the flux line near the twin plane.
In infinitely thick samples ($L\rightarrow\infty$) this binding free
energy is determined by the
zero-point energy of a fictitious two-dimensional quantum mechanical particle
confined to a one-dimensional potential well, or $U(T)=-E_0(T)$,
where $E_0(T)$ is the ground state eigenvalue of a two-dimensional
``Schr\"odinger'' equation,
\eqn\schrod{\Big[-{T^2\over 2\epst}\nabla^2_{\perp}+V_1(y)\Big]\psi_0(x,y)
=E_0\psi_0(x,y),}
and the twin plane is centered at $y=0$, i.e., $V_1(y)=-U_0$ for $|y|\leq b_0$,
$V_1(y)=0$ for $|y|>b_0$.
In the quantum mechanical analogy $T$ plays the role of Planck's constant
$\hbar$, $\epst$ that of the mass $m$ of the fictitious particle
and $L^{-1}$ that of the particle's temperature.
The $x$ and $y$ degrees of freedom are decoupled in Eq. \schrod\ and
the ground state wavefunction is
the product of a free-particle wavefunction in the $x$ direction
and the ground state wavefunction $\phi_0(y)$ of a one-dimensional
particle in a well \nref\landau{See, for instance, L.D. Landau and E.M.
Lifschitz, {\it Quantum Mechanics} (Pergamon, New York, 1965), Sec. 45.}
\refs{\landau},
$\psi_0 (x,y)={1\over\sqrt{D}}e^{iq_xx}\phi_0(y)$
\refs{\drnrev ,\nvlong}. Here $D$ is the system size in the $x$ direction
and $q_x$ the wavevector. The corresponding ground state energy is
$E_0(T)={T^2q_x^2\over 2\epst}+E_{0w}$.
The first term is the free particle contribution
describing the energy cost associated with localizing a flux line
within a distance $\sim 2\pi/q_x$ and $E_{0w}<0$ is the ground state
energy of a one-dimensional particle in a well \refs{\landau}.
If no other flux lines are present and the sample is infinite in the
$x$ direction, we can take $q_x$=0 and
$U(T)= -E_{0w}=U_0f(T/T^*)$ \refs{\drnrev}.
Here $T^*=\sqrt{2U_0\epst} b_0$ is a crossover temperature
above which thermal fluctuations delocalize the flux line, and
$f(x)\approx 1-{\pi^2\over 4}x^2$ for $x<<1$ and
$f(x)\approx 1/x^2$ for $x>>1$.
The probability of finding a point on the vortex at a transverse
displacement $\rpvec$ is proportional to $|\psi_0(\rpvec)|^2$ and
depends only on the transverse displacement $y$ relative to the center
of the twin plane. This corresponds to the fact that the flux line is
``free'' and therefore completely delocalized in the direction
parallel to the twin plane ($x$), while it is localized by the pinning
potential in the direction transverse to the twin
plane ($y$). The corresponding transverse localization length $l_y(T)$
can be defined as
\eqn\localiz{[2l_y(T)]^2=\int_{-\infty}^{+\infty}dy y^2 |\phi_0(y)|^2,}
where the wavefunction is assumed to be normalized.
As shown in \refs{\drnrev}, one finds
$l_{y}(T)\approx b_0[1+{\cal O}(T/T^*)]$, for
$T<<T^*$. At high temperature thermal wandering is important and
the localization length can become larger than $b_0$, with
$l_{y}(T)\approx {T\over\sqrt{2\epst U(T)}}\sim b_0(T/T^*)^2$, for
$T>>T^*$.

When many vortices are present, the repulsive intervortex interaction tends
to confine each flux line to a ``cage'' provided by the surrounding vortices
in a triangular lattice \nref\drnnato{D.R. Nelson, in Proceedings of the
NATO Advanced Study Institute on {\it Phase Transitions and Relaxation
in Systems with Competing Energy Scales}, T. Riste and D. Sherrington, eds.
(Kluwer Academic Pub., The Netherlands, 1993).}\refs{\drnnato}.
For fields below the ``filling'' field
$B_f\approx \phi_0/d^2$, defined as the field where
the flux lines fill the twin planes, forming a triangular
lattice of spacing $\sqrt{3}d/2$ \nref\notea{We neglect in this estimate
the anisotropy of the lattice arising from the pinning of vortices by the
twin planes. At low fields the difference between the intervortex spacing
$a_0$ in twin-free regions and the intervortex spacing $a_0^{TB}$
along the twin can be estimated from the magnetic flux decoration images
by Dolan et al. \refs{\dolan}, with the result $a_0^{TB}\approx 0.6 a_0$.}
\refs{\notea},
the additional confining potential provided by the repulsive interaction
does not change qualitatively the fluxon states in the direction transverse
to the twins, since in this direction flux lines are also localized by
the pinning potential. Interactions with neighbors do, however,
change qualitatively the behavior along the $x$ direction.
Following Ref. \refs{\drnnato}, a simple description of the role of
interactions can be obtained by considering
a representative fluxon localized near a single twin plane centered at
$y=0$ and subject to the additional confining potential provided by the
surrounding
vortices. If the position of all the other vortices is assumed to be fixed,
the confining potential can be approximated by a one-body effective
potential $V_{eff}(\rvec_1(z))$,
\eqn\veffect{V_{eff}(\rvec_1(z))={1\over N}\sum_{j\not= 1}
        V(|\rvec_1(z)-\rvec_j^0|),}
where $V(r)$ is the pair potential given in Eq. \pairint\ and
$\rvec_1(z)$ denotes the position of the representative fluxon.
The sum is over all the other vortices that are fixed at their equilibrium
positions $\rvec_j^0$, corresponding to the sites of the triangular lattice.
The free energy of the representative fluxon is then given by,
\eqn\harmonic{F_1^{eff}=\int_0^L dz \bigg[{\epst\over 2}
  \Big|{d\rvec_1(z)\over dz}\Big|^2
  +V_{eff}(\rvec_1(z))+V_1(|y_1(z)|)\bigg],}
where $V_1(|y|)$ is the single-twin pinning potential discussed earlier.
If we expand the effective potential $V_{eff}(\rvec_1)$ about its minimum
at $\rvec_1=0$, we find
\eqn\effpot{V_{eff}(\rvec_1)\approx V_{eff}(0)+{1\over 2}Cr_1^2,}
where, neglecting logarithmic corrections and constants of order unity,
\eqn\forcec{C\approx {2\epsilon_0\over a_0^2},}
for $\lambda_{ab}>>a_0$, where the pair interaction
$V(r)$ is logarithmic
($K_0(x)\approx -\ln x$, for $x<<1$), and
\eqn\forcecl{C\approx {2\epsilon_0\over\lp^2}\sqrt{\pi\lp\over 2a_0}
  e^{-a_0/\lp},}
for $\lambda_{ab}<<a_0$, where the pair interaction decreases
exponentially with distance.
Again, following Ref. \refs{\drnnato},
in the limit $L\rightarrow\infty$ the
partition function of this representative fluxon is written
in terms of the ground state eigenfunction and eigenvalue of
the ``Hamiltonian'' operator of a fictitious
quantum mechanical particle. Dropping the constant term in Eq. \effpot ,
the corresponding ``Schr\"odinger'' equation
is given by
\eqn\schrodho{\Big[-{T^2\over 2\epst}\nabla^2_{1}
+{1\over 2}Cr_1^2+V_1(|y_1|)\Big]\Psi_0(\rvec_1)=E_0\Psi_0(\rvec_1).}
The $x$ and $y$ degrees of freedom are decoupled and the
``Schr\"odinger'' equation \schrodho\ can be separated into two
one dimensional equations (to simplify the notation, we drop the subscript
$1$ on the location of the representative fluxon),
\eqn\schrodhox{\Big[-{T^2\over 2\epst}{d^2\over dx^2}
+{1\over 2}Cx^2\Big]g_x(x)=E_xg_x(x),}
and
\eqn\schrodhoy{\Big[-{T^2\over 2\epst}{d^2\over dy^2}
+{1\over 2}Cy^2+V_1(|y|)\Big]g_y(y)=E_yg_y(y),}
with $\Psi_0(\rvec)=g_x(x)g_y(y)$ and $E_0=E_x+E_y$.
In the $x$ direction the vortex line is described by the ground state
of a one-dimensional harmonic oscillator of frequency
$\omega_0=\sqrt{C/\epst}\approx(1/a_0)\sqrt{2\epsilon_0/\epst}$.
The ground state energy is $E_x=T\omega_0$ and the corresponding eigenfunction
is
\eqn\harmox{f(x)={1\over(\sqrt{2\pi}x^*)^{1/2}}e^{-(x/2x^*)^2},}
where $x^*=(T^2/2\epst C)^{1/4}$ is the characteristic length scale for
vortex fluctuations along the $x$ direction.
In the absence of the twin plane, the ground state in the $y$ direction
is also that of a harmonic oscillator of frequency $\omega_0$ and
the vortex is confined by interactions
within a region of radius $r^*=\sqrt{x^{*2}+y^{*2}}$, with $y^{*}=x^{*}$,
centered at its equilibrium position, $\rvec=0$.
The presence of the twin boundary modifies the potential in the
$y$ direction, leading to an
additional square well near the center of the harmonic potential, as
in Eq. \schrodhoy .
The range of the wavefunction $g(y)$ controls the localization
length $\loc$ in the direction transverse to the twin plane.
This is determined  by the interplay of
the length scale $y^*(T)$ for harmonic fluctuations and
the localization length $l_y(T)$ defined in Eq. \localiz\
associated with the pinning potential. These two length scales
are sketched in Fig. 5 as functions of temperature.
For $T<<T^*$, $l_y\approx b_0$. If the temperature is so low that
$y^*<l_y\approx b_0$,
the range of the wavefunction $g(y)$ is controlled
by interactions and $\loc\approx y^*$.
The characteristic temperature $T_{x1}$ where $y^*=b_0$
is given by $T_{x1}=(b_0/a_0)\sqrt{2\epsilon_0/U_0}T^*<<T^*$, as
shown in Fig. 5. For $T>>T^*$, $l_y(T)\approx b_0(T/T^*)^2$ grows
more quickly than $y^*$ with temperature. There is therefore a second crossover
temperature $T_{x2}$, as shown schematically in Fig. 5.
For $T>T_{x2}$, $l_y>y^*$ and the vortex line is confined
only by the harmonic well from intervortex interactions.
A lower bound for $T_{x2}$ can be obtained from
$y^*(T_{x2})\approx l_y\approx b_0(T_{x2}/T^*)^2$,
with the result $T_{x2}=[(a_0/b_0)\sqrt{U_0/2\epsilon_0}]^{1/3} T^*>T^*$.
For the parameters of interest here $T_{x2}$ is somewhat smaller than
the clean lattice
melting temperature $T_m$, defined by $y^*(T_m)\approx c_L a_0$, with
$c_L\approx 0.15-0.3$ the Lindeman constant, and is comparable to
the isolated vortex depinning temperature,
defined by $l_y(T_{dp})\approx d$. In this paper we only consider
the situation where all vortices are pinned in the ground state
and $T<<\min(T_{dp}, T_m)$.
We therefore restrict ourselves to $T<T_{x2}$ and
we then find $\loc\approx y^*$ for $T<T_{x1}$  and
$\loc\approx l_y$ for $T_{x1}<T<T_{x2}$.
In short, the ground state of a single vortex confined by the pinning potential
of a twin plane along the $y$ direction and by the isotropic ``cage''
provided by the repulsive interaction with the other vortices
is localized in all directions, with localization lengths
$\loc\approx l_y$ in the direction
transverse to the twin (for $T_{x1}<T<T_{x2}$) and
$l_{\parallel}\approx x^*$ in the direction parallel
to the twin plane.
The total binding free energy renormalized by interactions
is of order $U_R(T)\approx U(T)-T\omega_0$. For $T<<T_{x2}$
the harmonic oscillator zero point energy is always negligible
compared to the pinning energy $U(T)$ and $U_R(T)\approx U(T)$.

In the presence of a family of parallel twin planes the flux line
can ``tunnel'' between different localized states \refs{\drnnato , \nvlong}.
In this paper we are
interested in studying the response of a flux array pinned
by a family of parallel twin planes to a Lorentz force normal to the
twin planes for $B<<B_f$. In the ground state the flux lines
are all localized on the attractive twin planes.
The transverse driving force promotes motion of the vortices
between different twin planes, corresponding to ``tunneling''  between
different localized states along the $y$ direction,
while the repulsive interaction
confines the vortices in the direction parallel to the twin planes.
At low temperature the flux lines will move along the direction of
the driving force ($y$ direction) within one-dimensional channels
of width $\sim 2x^*$.
Using elementary quantum mechanics it can be shown \refs{\drnrev , \drnnato}
that the rate of tunneling between localized states on different
twin planes separated by a distance $d_{ij}$ is
$t_{ij}\sim 2 U(T)e^{-E_{ij}/T}$,
with $E_{ij}=\sqrt{2\epst U(T)}d_{ij}$. The energy $E_{ij}$ is the energy of
a ``kink'' configuration shown in Fig. 3b, connecting two pins at a
distance $d_{ij}$.

To study the low-lying excitations from this ground state arising from thermal
fluctuations one needs to sum over vortex trajectories by evaluating
appropriate path integrals.
As discussed in \refs{\nvlong} and \refs{\drnrev}, these configuration
sums closely resemble the imaginary time path integral formulation
of quantum mechanics of {\it two-dimensional} particles in a static random
potential $V_D(y)$. Many relevant results
regarding the statistical mechanics of flux lines can then be obtained from
elementary quantum mechanics.

The dynamics of flux lines
driven by a Lorentz force transverse to the twin planes
can then be described by a tight-binding model for {\it one-dimensional}
bosons \refs{\nvlong}. The lattice sites in the model are defined
by the $M$ positions $\{Y_i\}$ of the twin planes and the tight binding
Hamiltonian governing the dynamics in each one-dimensional channel is given by
\eqn\tight{H=-[\mu+U(T)]\sum_ia_i^{\dag} a_i +\sum_{i\not= j}t_{ij}
  (a_i^{\dag} a_j+a_j^{\dag} a_i)+ {V_0\over 2}\sum_i
  a_i^{\dag} a_i(a_i^{\dag} a_i-1).}
Here $\mu\approx\phi_0( H-H_{c1})/4\pi$
is the chemical potential which
fixes the flux line density, $a_i^{\dag}$ and $a_i$ are boson creation
and annihilation operators at site $Y_i$,
$t_{ij}$ is a tunneling matrix element connecting localized states
$i$ and $j$ and $V_0$ represents a typical energy cost for double
occupancy of a site of the one-dimensional tight-binding lattice.
As flux lines move in the transverse direction along the one-dimensional
channels, the repulsive intervortex interaction provides an energy cost
for an additional flux line occupying an already filled
twin.
The corresponding on-site repulsion $V_0$ can be estimated as
\eqn\interact{\eqalign{V_0&\approx V(b_0)-V(d) \cr
    &\approx 2\epsilon_0\big[\ln(\lp/\xi_{ab})-
     K_0(d/\lp)\big],}}
where $V(r)$ is the pair interaction given in Eq. \pairint\ and
we assumed $b_0\approx\xi_{ab}$.
If $d>>\lp$, we find $V_0\approx 2\epsilon_0\ln(\lp/\xi_{ab})$,
while for $d<<\lp$, we obtain $V_0\approx 2\epsilon_0\ln(d/\xi_{ab})$.
This estimate assumes that the various one-dimensional channels are completely
decoupled.

The first two terms of the tight-binding Hamiltonian
determine a noninteracting density of
states $g(\epsilon)$ (here $\epsilon$ is an energy per unit length
and $g(\epsilon)$ has units of $1/$energy), such that
${\cal N}(\epsilon)=\int_{-\infty}^{\epsilon}g(\epsilon')d\epsilon'$
is the number of localized states per unit length with energy less than
$\epsilon$. Note that $g(\epsilon)$ is normalized so that
${\cal N}(+\infty)=1/d$. Even if the pinning sites are all identical
in size and well depth, dispersion of energy levels arises because
vortices can tunnel between nearby twin planes.
The width $\gamma$ of the impurity band should then be of order
$\gamma\approx t(d)$, where $t(d)$ is the tunneling matrix element
evaluated at a typical twin spacing $d$. Interactions will further broaden the
band and one can estimate,
\eqn\bandw{\gamma\approx \max\{t(d), V_0\}.}
This bandwidth is practically always dominated by $V_0$. In particular
for the case $d<<\lp$ one finds
$\gamma\approx V_0\approx 2\epsilon_0 \ln(d/\xi_{ab})$ for all temperatures
$T<T^*d/b_0$.
If the localized states are filled up to a chemical potential $\mu$ such that
about half of the twins are occupied by at least one vortex, we can
approximate the density of states $g(\mu)$ corresponding to the
most weakly bound flux lines with energy $\epsilon\sim\mu$ as
$g(\mu)\approx 1/d\gamma$, i.e.,
\eqn\dos{dg(\mu)\approx\min\{1/t(d),1/ V_0\}.}
{}From our discussion of the bandwidth $\gamma$ we find that
the second term generally dominates for all temperatures of interest.
Then $g(\mu)$ is approximately temperature independent, with
$dg(\mu)\approx 1/V_0$.

\newsec{Vortex Dynamics at Low Temperatures}

We consider vortex transport in the
presence of a driving current ${\bf J}\perp{\bf H}$ parallel to the twin
planes, i.e., ${\bf J}=-J\xhat$. The applied current exerts a Lorentz
force per unit length on the vortices (see Fig. 1a),
\eqn\lorentz{{\bf f}_L={\phi_0\over c}\zhat\times{\bf J}=
           \yhat f_L,}
with $f_L=\phi_0J/c$,
and drives the vortices to move in the direction transverse to the twin planes,
leading to an additional term,
\eqn\freelor{\delta F_1=-f_L\int_0^L y_1(z) dz,}
in the single-vortex free energy, Eq. \freeone . In the context of the analogy
with boson quantum mechanics,
this term represents a fictitious ``electric field''
${\bf E}={1\over c}\zhat\times{\bf J}=\yhat J/c$ acting on particles with
``charge'' $\phi_0$.
The correspondence between the problem of carrier dynamics in disordered
semiconductors and vortex dynamics in the presence of correlated linear
or planar disorder is summarized in Table 2.

Up to numerical constants and logarithmic corrections, the critical current
at low temperatures can be obtained by equating the Lorentz force
to $U_0/b_0$, with the result
$J_c(0)\approx cU_0/\phi_0b_0$ \refs{\nvlong}.
Thermal fluctuations renormalize the critical current and one can
estimate $J_c(T)\approx cU(T)/\phi_0l_{\perp}(T)$. For $T>>T^*$,
$J_c(T)\approx J_c(0)(T^*/T)^4$.
The crossover temperature $T^*$ is itself a function of temperature.
We can define the temperature $T_1$ above which the entropy from
flux-line wandering is important in renormalizing the binding
free energy $U(T)$ by the self-consistency relation
$T^*(T_1)=T_1$\refs{\drnrev}.
At low temperature the interaction between a vortex line and a twin plane
is always attractive and $U_0\approx\alpha_b\epsilon_0\tau$,
where $\alpha_b<1$ is a dimensionless parameter related to the barrier
transparency and $\tau=1-T/T_c$\refs{\gesh}.
Using a mean field parametrization
of the critical fields, we find
$T^*(T)/T_c=\sqrt{{\alpha_b\ln\kappa\over 4Gi}}\tau$, where
$\kappa=\lambda_{ab}/
\xi_{ab}$ and $Gi=(\lambda_c^2/2\lambda_{ab}^2)(T_c/H_{c0}^2\xi_{ab0}^3)^2$
is the Ginzburg number, with $H_{c0}$ the
thermodynamic critical field at $T=0$ and $\xi_{ab}=\xi^0_{ab}\tau^{-1/2}$.
Using $\kappa\approx 10^2$, $Gi\approx 10^{-2}$ and $\alpha_b\approx 0.1$
\nref\decor{Decoration experiments in twinned samples at low temperatures
\refs{\dolan}
indicate that $U_0\sim 0.1\epsilon_0$ \refs{\blatter}.}\refs{\decor}
for YBCO, we find $T_1\approx 0.77 T_c$.

At low temperatures and fields well below $B_f$,
vortex dynamics is determined by the competition between pinning by the
one-dimensional array of twin planes and thermal fluctuations of the vortices.
In analogy with the case of columnar pins that was discussed in detail in
\nvlong , the boson mapping reduces single
vortex dynamics to a problem of hopping conductivity of localized
particles in one dimension.
The current density in the usual hopping conductivity problem corresponds to
the vortex velocity (i.e., voltage) and the electrical conductivity
maps onto the resistivity from vortex motion (see Table 2).
The low temperature dynamics
of vortices driven transverse to an array of parallel twin planes
presents the same
rich variety of hopping conductivity phenomena that occur in semiconductors,
as pointed out by Nelson and Vinokur for the case of columnar
pins \refs{\nvlong}. What is new here is that
vortex dynamics maps onto the problem of hopping conductivity in
one dimension. In this reduced dimensionality rare events, such as large
regions voids of twin planes, can dominate the transport at low currents
leading to new mesoscopic phenomena, as discussed in Section 5.

Here we are interested in the low temperature regime where
transport is dominated by single-vortex dynamics.
In this case the dominant contribution to dissipation can be described
in terms of the low-lying excitations from the ground state
that correspond to thermally activated jumps of vortex lines over the
relevant pinning energy barriers. The resistivity takes the
form given in Eq. \taff .
In the following we determine the barrier heights ${\cal U}(L,J)$ corresponding
to various transport regimes and the boundaries between the
various regimes in the $(L,J)$ plane.
In samples of finite thickness $L$ in the field direction the typical pinning
energy barriers ${\cal U}(L)$ grow with $L$ but are independent of current,
yielding a linear resistivity. In thick samples there is a nonlinear
resistivity associated with barriers ${\cal U}(J)$ that grow at low currents.
We assume that in the ground state all the flux lines are localized on
twin planes. We then study the low lying excitations from the ground
state that can be nucleated by a finite temperature $T$ or by a
driving current $J$. The largest contribution to the resistivity
from each class of excitations is assumed to be inversely proportional to the
shortest time for the nucleation of a given excitation. The latter is
determined by
the typical energy barrier ${\cal U}$ for the formation of the low-lying
excitations, which is identified with the
saddle point in the single-flux line free energy.
The discussion in this section follows closely that of Ref. \refs{\nvlong},
where the corresponding results for votices pinned by columnar defects
were obtained.
\bigskip

\noindent{\it Linear response}

Consider a fluctuation that extends a length $z$ along the twin and
a distance $y$ in the direction of the Lorentz force.
The free energy of this fluctuation relative to the case $f_L=0$ is
\eqn\freefluct{\delta F(y,z)\approx \epst {y^2\over z}
   +Uz-f_Lyz.}
Optimization of Eq. \freefluct\ with repsect to $z$ for $f_L=0$ yields the
shape of the optimal low temperatute fluctuation,
\eqn\optshape{y\sim\sqrt{U/\epst}~z.}
At low currents in samples of very small thickness $L$ there is
a linear resistivity due to the flow of rigid
flux-line segments of length $L$ and typical transverse width
$y_{rf}\approx \sqrt{U(T)/\epst}L$ obtained by letting $z\sim L$
in Eq. \freefluct .
The corresponding saddle point energy is ${\cal U}_{rf}(L)\sim U(T)L$,
resulting in a
linear ``rigid-flow'' resistivity,
\eqn\linrho{\rho_{rf}(L)\approx\rho_0e^{-U(T)L/T}.}
At larger currents
the contribution from the Lorentz force
to the single-line free energy \freefluct\
becomes comparable to the typical energy
barrier ${\cal U}_{rf}$. When $f_Ly_{rf}L>{\cal U}_{rf}$
or $J>J_L=c\sqrt{\epst U}/(\phi_0L)$, the response becomes nonlinear.
In the thermodynamic limit $J_L\rightarrow 0$ and the IV characteristic
is nonlinear at all currents.
The characteristic current $J_L$ is also conveniently expressed in
terms of the energy of a kink configuration connecting neighboring pins
separated by the distance $d$ (see Fig. 3a). The typical thickness $w_k$ of
a kink along the $z$ direction is obtained from Eq. \optshape\
for $y\sim d$, with the result $w_k=d\sqrt{\epst/U}$. The kink energy is
the corresponding saddle-point free energy,
$E_k=w_kU=\sqrt{\epst U(T)}d$. The energy barrier associated with the
``rigid-flow'' resistivity can then be written as
${\cal U}_{rf}=E_k(L/w_k)$ and the current scale for nonlinear
transport is $J_L=cE_k/(\phi_0Ld)$.

The line $J=J_L(L)$ defines the boundary in the $(L,J)$ plane that separates
the regions of linear ($J<J_L(L)$) and nonlinear ($J>J_L(L)$) response
(see Fig. 2).
The details of the $(L,J)$ phase diagram
are controlled by the dimensionless parameter $\alpha=g(\mu)dU(T)$.
Typical phase diagrams for $\alpha<E_k/T$ are shown in Fig. 2.
The rigid flow mechanism dominates the linear resistivity only in very
thin samples. When $w_k<L<L_1$, where
$L_1=E_k/\gamma$ is the length below which dispersion from
tunneling and interactions can be neglected,
transport occurs via the
hopping of vortices between nearest neighbor (nn)
pinning sites. This region of the phase diagram is only
present if $L_1> d\sqrt{\epst/U}$, or $\gamma <U$. For
$L>L_1$ dispersion is always important and the relevant
excitations are
superkinks (Fig. 3c), which correspond to
the tunneling of vortices between
remote pinning sites analogue to Mott's electronic conductivity
in disordered semiconductors. For $J>J_L(L)$ the resistivity is nonlinear.
At large currents the typical transverse displacement is smaller than the
average spacing $d$ between twin planes
and transport is dominated by ``half-loop'' excitations (Fig. 3a),
characterized
by an energy barrier that grows linearly as the current decreases,
${\cal U}_{hl}\sim 1/J$. Finally, at the smaller current flux motion
takes place via VRH, characterized by a diverging energy barrier,
${\cal U}_{VRH}\sim 1/J^{1/2}$.
We now discuss in more detail the origin of the various contributions
summarized in Table 1 and the estimate of the energy barriers.

We first consider the linear portion of the phase diagram ($J<J_L$)
in samples of increasing thickness $L$. When the typical transverse width
$y_{rf}\sim\sqrt{U/\epst)}L$ of a rigidly flowing flux segment becomes
comparable to $d$, transport occurs via nucleation of double kink
configurations (Fig. 3b) of energy $\sim 2E_k$.
The double kink then separates to $z=\pm\infty$, resulting in the hopping of
vortices from one pin to a neighboring one.
As discussed in \refs{\nvlong}, this transport mechanism
will dominate only if the sample is so thin that
the width $\gamma$ of the impurity band arising from tunneling
and interactions is negligible ($L<L_1=E_k/\gamma$). In this case flux motion
will occur via hopping between nearest neighbor pins,
resulting in a linear resistivity $\rho_{nnh}\sim \exp(-aE_k/T)$,
with $a$ a numerical constant.
In extremely thin samples this transport mechanism will ultimately
be suppressed. In fact for $L<w_k$ transport via the flow of rigid flux
segments described above is energetically favorable over nn hopping.
As a result,
a necessary condition for observing a linear
nearest neighbor hopping resistivity $\rho_{nnh}$ is
$L_1>w_k$, or $\gamma<U$. If we estimate the density of states
as $dg(\mu)\sim1/\gamma$, the condition $\gamma<U$ requires
$\alpha=g(\mu)dU>1$.
The Lorentz force term in Eq. \freefluct\ will modify the kink
energy and thickness. By optimizing Eq. \freefluct\ with respect to $z$
for $y\sim d$ and $f_L\not=0$, we find that a finite current increases
the thickness of a typical kink, according to
$\tilde{w}_k(J)=w_k(1-J/J_1)^{-1/2}$, where $J_1=cU/(\phi_0d)$.
This result only applies for $J<J_1$. At higher currents simple
nn hopping cannot occur.

In thick samples ($L>L_1$) the
dispersion of energies between different pinning sites makes motion
by nearest neighbor hopping energetically
unfavorable
(the energy barrier diverges with the sample thickness $L$). Tunneling
occurs instead via the formation of ``superkinks'' (Fig. 3c)
that throw a vortex
segment onto a spatially remote pin connecting
states which optimize the tunneling probability.
The free energy of a superkink excitation shown in Fig. 3c
relative to the case $f_L=0$ is then \refs{\nvlong , \hwa},
\eqn\freesk{\delta F_{sk}\approx 2E_k(y/d)+\Delta\epsilon z
- -f_Lyz.}
We assume all states up to a chemical potential $\mu$ are filled.
The states available to a weakly bound flux line about to hop a distance
$y$ are those within an energy $\Delta\epsilon$ determined by requiring
that there is at least one localized state within a region $(y,\Delta\epsilon)$
of configuration space, i.e.,
$g(\mu)y\Delta\epsilon\simeq 1$.
The shape of the most important superkink excitations is
obtained by minimizing Eq. \freesk\ for $f_L=0$ and is given by
\eqn\bestsk{y\sim\sqrt{dz\over E_kg(\mu)}.}
In finite thickness samples the saddle point free energy corresponding to
superkink fluctuations of width given by Eq. \bestsk\ for $z\sim L$ yields
a linear Mott resistivity,
given by
\eqn\reslm{\rho_{Mott}(L)\approx\rho_0 e^{-E_k(L/\alpha w_k)^{1/2}},}
with $\alpha=U(T)g(\mu)d$.
\bigskip

\noindent{\it Nonlinear response}

In the nonlinear regime ($J>J_L(L)$) the contribution to the free
energy from the Lorentz force cannot be neglected when estimating the energy
of the dominant excitations.
For $J_1<J<J_c$, with $J_1=cU(T)/\phi_0d$,
flux motion occurs via thermally
activated ``half-loop'' configurations identical to
those discussed in \nvlong\ for the case of columnar pins. The length and
width of an unbound line segment for the lowest-lying
half-loop excitations are obtained by minimizing the free energy \freefluct\
for $f_L\not= 0$, with the result
$z_{hl}\sim (U_0\epst)^{1/2}/f_L$
and $y_{hl}\sim (U_0/\epst)^{1/2}z_{hl}\sim U(T)/f_L$, respectively.
The saddle point
energy of a half-loop excitation is ${\cal U}_{hl}\approx \sqrt{\epst
U^3(T)}/f_L$,
yielding a nonlinear resistivity,
\eqn\ressl{\rho_{hl}\approx \rho_0\exp[-(E_k/T)(J_1/J)].}
In the context of the mapping of flux-line dynamics onto the
problem of hopping conductivity, the nucleation of half loops corresponds
to tunneling of a carrier from a localized state directly into conduction
band, as shown in Fig. 6.

For $J<J_1$ the size of the transverse displacement of the liberated vortex
segment exceeds the average distance $d$ between twin planes and transport
occurs via variable range hopping (VRH) which generalizes the Mott mechanism
to the nonlinear case. Again a flux line hops to a state within
a region $(y,\Delta\epsilon)$ of phase space, with
$g(\mu)y\Delta\epsilon\sim 1$.
The size of the most important excitations is determined by minimizing
Eq. \freesk\ with $f_L\not=0$, with the result
$y_{VRH}\sim (g(\mu)f_L)^{-1/2}$ and
$z_{VRH}\sim E_k/df_L$.
One then obtains a non-Ohmic VRH behavior with
\eqn\resvrh{\rho_{VRH}\approx\rho_0e^{-{\cal U}_{VRH}(J)/T}=
\rho_0\exp[-(E_k/T)(J_0/J)^{1/2}],}
with $J_0=J_1/\alpha$.
The crossover to the linear Mott resistivity
resistivity takes place when $z_{VRH}\sim L$, or $J\sim J_L$, consistent with
the
result obtained above when discussing half-loop excitations.
The VRH contribution to the resistivity dominates that from half loop only if
if ${\cal U}_{VRH}<{\cal U}_{hl}$, or $J<J_2=J_1\alpha$.

The above results are summarized in Table 1. The corresponding
phase diagrams are shown in Figs. 2 for $\alpha<E_k/T$.
There are three relevant current scales, $J_0=J_1/\alpha$, $J_1=cU/\phi_0d$
and $J_2=J_1\alpha$, all much smaller than the pair breaking current
$J_{pb}=4c\epsilon_0/(3\sqrt{3}\phi_0\xi_{ab})$.
For $\alpha>1$ one can have $L_1>w_k$ and there is a region of the phase
diagram where transport occurs via nn hopping (Fig. 2a).
For $\alpha<1$ nn hopping can occur only if
the chemical potential $\mu$ falls in the tails of the impurity band,
so that $g(\mu)d<1/\gamma$.
If $\mu$ falls well within the impurity band, so that $g(\mu)d\sim1/\gamma$,
then $\alpha<1$ requires $\gamma>U$ and nn hopping is always suppressed
in this case.
The Mott and the rigid flow
regimes are separated by a horizontal line above which
${\cal U}_{Mott}<{\cal U}_{rf}$. Similarly, the condition
${\cal U}_{VRH}={\cal U}_{hl}$
yields the vertical line separating the VRH and half loop regions.
\bigskip

\noindent{\it Collective effects}

At very low currents and in thick samples collective effects are
always important and flux motion takes place via the creep of
vortex bundles, rather than single vortices. The region where collective
effects dominate is shown schematically in Figs. 2 and 4. It corresponds
to the upper left portion of the $(L,J)$ phase diagram.
As discussed in Ref. \refs{\tora}, the crossover from
single vortex creep to creep of vortex bundles occurs when
\eqn\crossbund{L_z=a_0,}
where $L_z$ is the size of a typical single-vortex fluctuations
along the $z$ direction and $a_0$ the intervortex spacing.
The condition \crossbund\ is simply obtained by equating the
tilt energy $E_{tilt}$ of a single disorted vortex to the
elastic energy $E_{int}$ of interaction of with its neighbors.
The elastic energy associated with displacing a length $L_z$ of vortex line
at an average distance $a_0$ from its
neighbors a distance $u$ out of its
equilibrium position in the $xy$ plane is $E_{int}\sim c_{66}u^2L_z$,
where $c_{66}\sim\epsilon_0a_0^2$, and grows with $L_z$. In contrast,
the corresponding single-vortex
tilt energy, $E_{tilt}\sim\epsilon_0(u/L_z)^2L_z$, decreases as $L_z$
increases. Consequently when the longitudinal size of the typical fluctuation
is sufficiently large, or $L_z>a_0$, then $E_{int}>E_{tilt}$ and
collective effects are important.

In the VRH regime the relevant length scale is the width $w_{sk}$
of a superkink excitation, shown schematically in Fig. 3c, where
$w_{sk}\approx\sqrt{\tilde{\epsilon}_1/U}y_{sk}$, with $y_{sk}$
the typical size of a superkink in the direction of flux motion.
Collective effects dominate when $w_{sk}\geq a_0$.
In the linear regime ($J<J_L$) $y_{sk}$ is given by Eq. \bestsk\
with $z\sim L$, or $w_{sk}\approx\sqrt{w_kL}$, with $w_k=E_k/U$
the width of a kink (see Fig. 3d).
Dissipation is then dominated by creep of vortex bundles
for  $L\geq L_b=a_0^2/w_k$.
In the nonlinear regime ($J<J_L$) the size of the superkinks grows
with decreasing current and $w_{sk}\approx w_k(J_0/J)^{1/2}$.
Transport is always dominated by collective effects at sufficiently
low currents, i.e., for $J\leq J_b=J_0(w_k/a_0)^2$.
The crossover from single-vortex creep
to creep of vortex bundles is marked by the dashed lines
$L=L_b$ and $J=J_b$ in Figs. 2 and 4.
For $\alpha>1$ this crossover takes place well into the VRH region,
as shown in Fig. 2a, provided $L_b>L_1$ and $J_b<J_1$, which
corresponds to $B<(B_f/\alpha)(U/2\tilde{\epsilon}_1)$ (here and below we
assume the chemical potential falls in the middle of the impurity band and
$dg(\mu)\sim 1/\gamma$). For $\alpha<1$ (Fig. 2b) this crossover occurs
within the VRH region provided $L_b>w_k/\alpha$ and $J_b<J_2$, or
$B<B_f\alpha(U/2\tilde{\epsilon}_1)$

\newsec{Transport in the Presence of Tilt}

We now consider another transport geometry investigated in some of
the experiments by Kwok et al \refs{\kwoktwo}. Here the external field ${\bf
H}$ is tilted
at an angle $\theta$ away from the $c$ axis and out of the twin planes
(see Fig. 1b). The transport current is still applied along the twin planes,
which contain the $c$ axis, ${\bf J}=-\xhat J$, and the resulting Lorentz
force, ${\bf f}_L=(\phi_0/c)[\zhat\cos\theta+\yhat\sin\theta]\times{\bf J}$,
has components both normal to the twin planes and along the $c$ axis.
Only the $y$ component of the Lorentz force is effective at driving flux
motion
normal to the twin planes and therefore determines the voltage in the direction
of
the applied current.
The experiments by Kwok et al. \refs{\kwoktwo} have been mostly carried
out at high fields
for flux arrays in a liquid state, in a regime where intervortex interactions
are believed to be important. Here in contrast we neglect intervortex
interactions and investigate the dependence
of transverse transport on tilt angle in the regime where single-line
dynamics dominate.
Even though our result are therefore not directly relevant to
the experiments by the Argonne group, the strong angular dependence
that we predict for the resistivity is qualitatively similar to that reported
in the experiments.

The free energy of a fluctuation that extends a length $z$ along the twin
and a distance $y$ in the direction of average motion is
obtained by adding the tilt energy Eq. \freefluct , with the result,
\eqn\freetilt{\delta F(y,z,\theta)\approx \epst {y^2\over z}
   +Uz-f_L\cos\theta y~z -{\phi_0\over 4\pi}H_{\perp}y,}
where $H_{\perp}=H\sin\theta$ is the component of the field along the $y$
direction and $f_L=\phi_0J/c$, as in the preceeding sections.
Optimizing Eq. \freetilt\ with respect to $z$ for $f_L=0$, we find that
the shape of the optimal low temperature fluctuation is still given by
Eq. \optshape\ and does not depend on the angle $\theta$.
As shown by Hwa et al. \refs{\hwa}, the energy $\tilde{E}_k(\theta)$
of a kink fluctuation in the presence of tilt,
corresponding to the saddle point of the free energy \freetilt\
with $f_L=0$ for $y\sim d$ and $z\sim d\sqrt{U/\epst}$, is reduced
compared to its value for $\theta=0$, according to,
\eqn\kinktilt{\eqalign{\tilde{E}_k(\theta)&
            =E_k-{\phi_0H_{\perp}\over 4\pi}d \cr
           &=E_k\Big(1-{\sin\theta\over\sin\theta_c}\Big).}}
Here we have introduced a critical angle $\theta_c$ defined by
$\sin\theta_c=H/H_c$, with $H_c=4\pi E_k/\phi_0d$.
For $\theta>\theta_c$, the kink energy becomes negative and
kinks proliferate, as discussed in \refs{\hwa}. We now consider the angular
dependence of transport for $\theta<\theta_c$.

As in the case $\theta=0$, at high enough currents flux motion will occur
via the nucleation of half loops. By identifying the typical energy barrier
$\tilde{\cal U}_{hl}(\theta)$ for a half loop excitation in the presence
of tilt with
the saddle point energy found by minimizing Eq. \freetilt\ for $f_L\not=0$,
we obtain,
\eqn\hltilt{\tilde{\cal U}_{hl}(\theta)={{\cal U}_{hl}\over\cos\theta}
    \Big(1-a{\sin\theta\over\sin\theta_c}\Big),}
with $a$ a numerical constant of order one and ${\cal U}_{hl}=E_k(J/J_1)$
the half loop energy barrier for $\theta=0$.
The energy for nucleating a half loop excitation is reduced by the tilt.
The angular dependence of the resulting flux-flow resistivity
is very strong, since the angle appears in the argument of the exponential.
Flux motion will occur via half loop excitation provided the typical transverse
size of the half loop does not exceed the average distance between twin planes.
This imposes a lower bound on the values of the current where half loop
excitation dominates transport, given by $J>J_1(1+a'\sin\theta/\sin\theta_c)$,
with $a'$ a numerical constant of order one.
Tilt decreases the range of currents where half-loops dominate.

At lower currents transport will take place via VRH.
The angular dependence simply replaces the kink energy $E_k$
by the smaller kink energy $\tilde{E}_k(\theta)$ in the presence of tilt
given in Eq. \kinktilt .
Carrying then through the standard VRH argument described in the previous
section, one obtains a nonlinear angle-dependent resistivity given by
Eq. \taff , with
\eqn\vrhtilt{\tilde{\cal U}_{VRH}(\theta)=E_k(J_1/\alpha J)^{1/2}
        {1-\sin\theta/\sin\theta_c\over\sqrt{\cos\theta}}.}
Again the corresponding resistivity is a rapidly varying function of
angle, as observed in experiments. On the other hand, a simple estimate
using typical parameters
for $YBCO$ gives a very small value for the critical angle,
$\sin\theta_c\approx 0.1 H_{c1}/H$.
The transport experiments probe, however, linear
transport in the flux-liquid phase, where collective effects in the
flux-line dynamics are important.
The present dimensional analysis is useful in that it shows that
even in the regime of single-line dynamics, the presence of twin planes
naturally introduces a very sharp dependence of the resistivity
on tilt angle.

\newsec{Rare Fluctuations}

The results described in the Section 3
are qualitatively similar to those discussed
in \refs{\nvlong} for the case of flux arrays in the presence of
columnar pins. The most important difference for samples with parallel
arrays of twin planes is that
due to the one-dimensional nature of vortex transport
at low temperature, a new regime can arise at low current, where
flux-line dynamics is dominated by rare fluctuations
in the spatial distribution
of twin planes.
The vortex line can encounter a rare region where
no favorable twins are available at the distance of the optimal jump.
The vortex will then remain trapped in this region for a long time
and the resistivity can be greatly suppressed.
Rare fluctuations can also occur in samples with
columnar pins, but in that case because of the two-dimensional nature
of the problem, they will dominate transport and suppress
the resistivity
only at extremely small fields, when the number of rare regions exceeds
the number of vortices.

At a given temperature and for applied currents below $J_L$,
a vortex can jump from one twin plane
to another
at a distance $y$ only if the energy difference per unit length between the
initial and final configuration is within a range
$\Delta \epsilon\sim E_ky/Ld$.
A trap is then a region of configuration space $(y,\epsilon$)
void of localized states within a spatial distance $y$ and  an energy band
$\Delta\epsilon$ around
the initial vortex state. A vortex that has entered such a trap or ``break''
will remain in the trap for a time
$t_w\approx  t_0\exp(2y/\loc)$,
where $\loc$ is the transverse localization length and $t_0$
is a microscopic time scale.
The probability
of finding such a break is given by a Poisson distribution,
$P(y)\approx P_0(y)\exp[-Ag(\mu)y\Delta\epsilon]$,
where $P_0(y)$ is the concentration of localized states in the energy band
$\Delta\epsilon$, $P_0(y)\approx 2Ag(\mu)\Delta\epsilon$ and $A\sim 1$
is a numerical constant. The mean waiting time between jumps
is given by
\eqn\meant{\overline{t_w}\approx\int_0^{\infty} dy P(y)
t_0e^{2y/l_{\perp}(T)}.}
For $L>>L^*=\alpha w_k(T/E_k)^2$,
the integral can be evaluated at the saddle point, corresponding to the
situation where the mean waiting time is controlled by ``optimal breaks''
of transverse width $y_{l}^*\approx \loc L/L^*$,
with the result,
\eqn\opttime{\overline{t_w}\sim t_0 \sqrt{L^*/L}e^{L/L^*}.}
The optimal breaks are those that correspond to the longest trapping time
and will therefore be most effective at preventing flux motion and
dissipation.
The inverse of the trapping or waiting time determines the characteristic
rate of jumps, i.e., the velocity. The vortex velocity corresponding to the
optimal hopping rate  of Eq. \opttime\
yields a linear resistivity in finite-thickness samples, given by
\eqn\resbreakl{\rho_{bl}\approx\rho_0{T\over{\cal U}_{Mott}}
    e^{-({\cal U}_{Mott}/T)^2},}
where ${\cal U}_{Mott}$ is given in Table 1 and we have used
$L/L^*=({\cal U}_{Mott}/T)^2$.

For currents above $J_L$, the typical energy per unit length available
to a flux line for jumping a distance $y$ is $\Delta\epsilon\sim f_Ly$.
The corresponding nonlinear contribution to the resistivity from
traps of extent $(y, \Delta \epsilon)$ in configuration space
is again proportional to the inverse of the average waiting time
defined in Eq. \meant .
Again, for $\loc^2g(\mu)f_L<<1$ or $J<<J^*=\alpha J_1(E_k/T)^2$,
the integral can be evaluated at the saddle point, corresponding to an
optimal break width $y_b^*\approx [g(\mu)l_{\perp}f_L]^{-1}$,
with the result,
\eqn\resbreak{\rho_b\approx\rho_0{T\over{\cal U}_{VRH}}
   \exp[-({\cal U}_{VRH}/T)^2].}
It is clear by comparing Eqs. \ressl\ and \resvrh\ to
Eqs. \resbreakl\ and \resbreak , respectively, that the contribution
to the resistivity from tunneling \`a la Mott (both in the linear and nonlinear
regimes) would always dominate that from hopping between
rare optimal traps
if both mechanisms of transport can occur.
On the other hand,
in one dimension if the sample is wide enough in the direction of
flux-line motion to contain optimal traps, tunneling \`a la Mott
simply cannot take place because flux lines cannot get around the traps.
These rare traps with large waiting times will then control
the transport.
If $W$ is the sample width in the $y$ direction, the condition
for having optimal traps of width $y_{l,b}^*$ is $P(y_{l,b}^*)W>1$.
Optimal traps will therefore be present only if $J>J_w=J^*/\ln(2W/\loc)$
for $J>J_L$ and if $L<L_w=L^*\ln(2W/\loc)$ for $J<J_L$.

These are, however, only necessary conditions for the sample to contain
many optimal breaks. They do not guarantee that these breaks will
dominate transport. A flux line can in fact escape a break by nucleating
a half-loop excitation or, in the language of semiconductor transport,
by tunneling directly from a localized state into conduction band (see
Fig. 6).
This will occur if the transverse size of a typical half loop exceeds
the size of the optimal break, i.e., if $y_{hl}>y_b^*$, or
$\alpha<E_k/T$. In the context of the
analogy with boson quantum mechanics this condition translates into
the requirement that the spatial distance between the occupied localized state
and the conduction band edge in the presence of the applied current
is shorter than the size of the trap (see Fig. 6).
If the flux line can escape the trap by half-loop nucleation,
breaks will never dominate transport and their only effect
will be that of possibly suppressing VRH in a region of the phase diagram
and extending to lower currents the region where transport occurs
via half-loop nucleation.
This can occur if $J_w$ is smaller than the scale
setting the high current boundary of the VRH region, or
$J_w<\min(J_1,J_2)$.
For instance if $J_w<J_2$, or $\ln(2W/\loc)>(E_k/\alpha T)^2$, with $\alpha<1$,
rare fluctuations will modify the phase diagram of Fig. 2b
by pushing the high current boundary
of the VRH region down to $J_w$.
Similar considerations apply to the linear response.
On the other hand, if $y_{hl}<y_b^*$, or $\alpha>E_k/T$,
there will be a portion of the $(L,J)$ phase diagram where breaks dominate
transport, as shown in Fig. 4.

For YBCO, we estimate $E_k\sim 1K\AA^{-1} d$.
Assuming $\alpha\sim U/\gamma$, the condition $\alpha>E_k/T$
can only be satisfied at low fields ($B<1KG$ for $d\sim 200\AA$).
The sample will contain optimal breaks if $W>30\AA\exp(J^*/J)$,
with $J^*\sim 4\times 10^5$Amp/cm$^2$ at $80K$.

If the sample is too short to contain optimal breaks,
i.e., $W P(y\sim y^*_{l,b})<1$, the dynamics
is controlled  by the trap with the longest waiting time,
$t(y_f)\sim \exp(y_f/l_{\perp})$, with $y_f$
determined by the condition $W P(y_f)\sim 1$. The corresponding
resistivity is proportional to this smallest hopping rate,
\eqn\traprho{\rho_W\approx\rho_0e^{-y_f/l_{\perp}}.}
In this case the relevant physical quantity is the logarithm
of the resistivity,
\eqn\reslog{\ln(\rho_W/\rho_0)=-y_f/l_{\perp}
  \approx - {{\cal U}_{VRH}\over T}
  \Big\{\ln\Big[{2W\over l_{\perp}}{T\over {\cal U}_{VRH}}
     \Big(\ln(2W/ l_{\perp})
  \Big)^{1/2}\Big]\Big\}^{1/2}. }
The leading dependence of Eq. \reslog\ on current and temperature
is the same as that of the VRH contribution.
Equation \reslog\ also contains, however, logarithmic terms that
in sufficiently short samples will give a random spread of values
of the resistivity from sample to sample. These effects
have been discussed for semiconductors \refs{\ruzin}.
In this case a more relevant
physical quantity rather than the resistivity itself is the
distribution of the logarithms of the resistivity over different samples.
The expression \reslog\ determines the position of the maximum of
this distribution.

\vskip .2in
This work was supported by the National Science Foundation through
Grants No. DMR91-12330 and DMR92- 17284
and through the U.S. Department of Energy,
BES-Material Sciences, under contract No. W-31-109-ENG-38.
MCM is grateful to Argonne National Laboratory and to the Institute for
Scientific
Interchange in Torino, Italy, for hospitality and partial support
during the completion of this work.
VMV thanks Gianni Blatter and the Swiss
National Foundation for supporting his visit at ETH-Z\"urich were
part of this work was carried out.
Finally, we both thank David Nelson for many stimulating discussions.

\vfill\eject
\listrefs


\vfill\eject
\vskip .5in

\begintable
Linear|Nonlinear\crthick
${\cal U}_{rf}=UL=E_k(L/w_k)$| ${\cal U}_{hl}=E_k(J_1/J)$ \cr
${\cal U}_{nnh}=E_k(1-J/J_1)^{-1/2}$| \cr
${\cal U}_{Mott}=E_k(L/\alpha w_k)^{1/2}$| ${\cal U}_{VRH}=E_k(J_1/\alpha
J)^{1/2}$
\endtable

\vskip .2truein
\noindent Table 1. Energy barriers determining the various contributions to
the resistivity of Eq. \taff , with $\alpha=U(T)g(\mu)d$.

\vfill\eject

\begintable
CARRIERS|VORTICES\crthick
$m$| $\tilde{\epsilon}_1$ \cr
$\hbar$| $k_BT$ \cr
$\beta\hbar$ | $L$ \cr
single impurity level $E_{D}$($E_A$) | $U(T)$ \cr
$\mu$|$(\phi_0/4\pi)(H-H_{c1})$\cr
$\vec{E}$|${1\over c}\hat{z}\times\vec{J}$\cr
carrier velocity $\sim$ current density|vortex velocity $\sim $ voltage\cr
conductivity $\sigma$| resistivity $\rho$\cr
conduction-band transport| flux flow\cr
tunneling from impurity levels to conduction band| half-loop\cr
VRH|superkink
\endtable

\vskip .2truein
\noindent Table 2. Correspondence between carrier dynamics in disordered
semiconductors and vortex dynamics in the presence of correlated disorder.

\vfill\eject

\figures

\fig{1}{Geometry of the transport experiment corresponding to strong
pinning by twin boundaries. In (a) the external field is aligned
with the
$c$ axis and lies in the plane of the twins. In (b) the
external field is
tilted at an angle $\theta$ out of the plane of the twin. In this case
only the $y$ component $f_L\cos\theta$ of the Lorentz force is effective at
driving flux motion transverse to the twins.}

\fig{2}{The $(L,J)$ phase diagram for $\alpha=g(\mu)dU\approx U/\gamma<E_k/T$.
The curved phase boundaries between the Mott and VRH regimes and between
the rigid flow and half loop regimes are determined by $J_L=cE_k/\phi_0dL$.
For currents below the characteristic current scale $J_1$
the typical transverse size of a fluctuation
in the nonlinear portion of the diagram
exceeds the average distance $d$ between pins.
The corresponding
length scale $w_k=E_k/U$ is the width of a kink connecting
pins at the distance $d$.
The crossover from half-loop to VRH is determined by $\min(J_1,J_2)$, with
$J_2=\alpha J_1$.
Figure 2a is for $\alpha >1$, corresponding to $L_1>w_k$, where $L_1$ is the
sample thickness above which level dispersion is important.
In this case flux motion can take place via nn hopping for $w_k<L<L_1$
and $J<J_1$.
Figure 2b is for $\alpha <1$, when $L_1<w_k$ and nn hopping is
suppressed.
The typical energy barriers determining
the resistivity in the various regimes are given in Table 1. The dashed
lines in the upper left corner of the plane delimit the region where
collective effects are important (see text).}

\fig{3} {Schematic representation of the various low-lying excitations
discussed in the text: (a) half-loop excitation, (b) double-kink configuration,
with $w_k=d\sqrt{\epst/U}$,
and (c) double-superkink configuration required for VRH.}

\fig{4} {The $(L,J)$ phase diagram for $\alpha=g(\mu)dU>E_k/T$.
In this case there is a region where the TAFF resistivity is controlled
by rare regions, both above and below $J_L$. The width of this region
is controlled by the sample size $W$ in the direction of flux motion.}

\fig{5} {The localization lengths in the direction transverse to the twin
planes: $l_y(T)$ is determined by the pinning potential
of the twin and $y^*(T)$ is determined by intervortex
interactions.
The various temperature scales are discussed in the text.}

\fig{6} {Schematic skecth in configuration space $(\epsilon, y)$
illustrating that a half-loop excitation
corresponds to tunneling of
the fictitious quantum mechanical particle directly into conduction band.
The straight line of slope $-f_L/U(T)$ is the conduction band edge in the
presence
of the fictitious electric field due to the Lorentz force.
A carrier occupying a localized state at $y=0$ near the center of the
impurity band, i.e., at an energy $\sim U(T)$ below conduction band edge,
is brought directly into conduction band by a hop of transverse
size $y_{hl}\approx U(T)/f_L$.}

\end